\def\wgt{\mathop{\rm wgt}}

\def\rank{\mathop{\rm rank}}

\def\diag{\mathop{\rm diag}}
\def\css{\mathop{\rm CSS}}
\def\gb{\mathop{\rm GB}\nolimits}

\def\LL{\mathop{\mathrm L}}
\def\lp{\mathop{\rm LP}}
\def\RR{\mathop{\mathrm R}}
\let\onlinecite\cite
\documentclass[10pt,conference,letterpaper]{IEEEtran}

\usepackage[T1]{fontenc} 
\usepackage[colorlinks,citecolor=blue,driverfallback=pdftex]{hyperref}
\usepackage{url}
\usepackage{amsmath}
\usepackage{amssymb}
\usepackage{xcolor}

\usepackage{soul} % strikeout, underline, highlight
\setulcolor{blue}
\setstcolor{red}
\definecolor{lightblue}{rgb}{.90,.95,1}
\sethlcolor{lightblue}

\usepackage{amsfonts}
\usepackage{graphicx}
\graphicspath{{./}{./figs/}}

\usepackage[inline]{enumitem}
\setlist{nosep}

\usepackage{amsthm}
\newtheorem{theorem}{Theorem}
\newtheorem{statement}[theorem]{Statement}

\newtheorem{example}[theorem]{Example}

\def\bs#1{\boldsymbol{#1}}

\usepackage{cite}

\begin{document}

\title{Abelian and non-abelian quantum two-block codes}

\author{\IEEEauthorblockN{Renyu Wang, Hsiang-Ku Lin, and Leonid P. Pryadko\thanks{leonid.pryadko@ucr.edu}}
    \IEEEauthorblockA{Department of Physics \& Astronomy, University of
  California, Riverside, California 92521 USA}
}
%%% \author{Renyu Wang}

%%% \affiliation{Department of Physics \& Astronomy, University of
%%%   California, Riveside, California 92521 USA}

%%% \author{Hsiang-Ku Lin}

%%% \affiliation{Department of Physics \& Astronomy, University of
%%%   California, Riveside, California 92521 USA}

%%% \author{Leonid P. Pryadko}
%%% \email{leonid.pryadko@ucr.edu}
%%% \affiliation{Department of Physics \& Astronomy, University of
%%%   California, Riveside, California 92521 USA}
%   \date{\today}
%\IEEEspecialpapernotice{(\today)}
\maketitle

\begin{abstract}
  We discuss quantum two-block codes, a large class of CSS codes
  constructed from two commuting square matrices.  Interesting
  families of such codes are generalized-bicycle (GB) codes and
  two-block group-algebra (2BGA) codes, where a cyclic group is
  replaced with an arbitrary finite group, generally non-abelian.  We
  present code construction and give several expressions for code
  dimension, applicable depending on whether the constituent group is
  cyclic, abelian, or non-abelian.  This gives a simple criterion for
  an essentially non-abelian 2BGA code guaranteed not to be
  permutation-equivalent to such a code based on an abelian group.  We
  also give a lower bound on the distance which, in particular,
  applies to the case when a 2BGA code reduces to a hypergraph-product
  code constructed from a pair of classical group codes.
\end{abstract}

\begin{IEEEkeywords}
  CSS codes, QECC, quantum LDPC codes, group algebra codes, group
  codes, two-block codes, 2BGA codes, GB codes, generalized bicycle codes
\end{IEEEkeywords}

\section{Introduction}
\label{sec:introduction}
\IEEEPARstart{G}{enerally,} any family of quantum low-density
parity-check (LDPC) codes with stabilizer generators of bounded weight
and distance scaling logarithmically or faster with the block length
has a finite fault-tolerant threshold to scalable error
correction\cite{Dennis-Kitaev-Landahl-Preskill-2002,%
  Kovalev-Pryadko-FT-2013,Gottesman-overhead-2014,%
  Dumer-Kovalev-Pryadko-bnd-2015}.  Recently, there was a significant
progress in constructing such codes\cite{Panteleev-Kalachev-2019,%
  Evra-Kaufman-Zemor-2020,Kaufman-Tessler-2021,%
  Hastings-Haah-ODonnell-2020,Panteleev-Kalachev-2020,%
  Breuckmann-Eberhardt-2020,Panteleev-Kalachev-2021}.  Unfortunately,
many of the proposed ``product'' constructions, e.g., in
Refs.~\onlinecite{Tillich-Zemor-2009,Bravyi-Hastings-2013,Zeng-Pryadko-2018,%
  Zeng-Pryadko-hprod-2020,%*Zeng-Pryadko-erratum-2022,%
  Hastings-Haah-ODonnell-2020,%
  Panteleev-Kalachev-2020,Breuckmann-Eberhardt-2020,Panteleev-Kalachev-2021},
tend to give rather long codes, and the existing lower bound for the
generator weight to give asymptotically good quantum LDPC codes with
finite rates and linear distance scaling is also very
large\cite{Panteleev-Kalachev-2021}.

In comparison, much shorter quantum codes, including quantum LDPC
codes with bounded generator weights, can be constructed with a
two-block anzats\cite{Kovalev-Pryadko-Hyperbicycle-2013}, a
construction based on a pair of square commuting matrices.  It gives a
family of Calderbank-Shor-Steane (CSS)
codes\cite{Calderbank-Shor-1996,Steane-1996} with relatively small
block lengths, twice the size of the original matrices.  The
commutativity can be achieved, e.g., by taking a pair of circulant
matrices, which gives generalized bicycle (GB)
codes\cite{MacKay-Mitchison-McFadden-2004,Kovalev-Pryadko-Hyperbicycle-2013,%
  Panteleev-Kalachev-2019,Wang-Pryadko-2022}, or using an arbitrary
finite abelian group instead of the cyclic
group\cite{Kalachev-Panteleev-2020}.  An important advantage of
two-block quantum LDPC codes is an overcomplete set of minimum-weight
stabilizer generators which may improve their performance in the
fault-tolerant setting.  Finally, GB and more general two-block codes
include certain families of hypergraph-product (HP)
codes\cite{Tillich-Zemor-2009} as a subclass, which guarantees the
existence of finite-rate codes with ${\cal O}(\sqrt n)$ distance
scaling in this family, but they also include codes with linear
distances\cite{Wang-Pryadko-2022}.  In comparison, the distance of an
HP code with the block length $n$ cannot exceed $\sqrt n$.

In this paper we discuss general quantum two-block codes.  We
introduce a family of \emph{two-block group algebra} (2BGA) codes
based on an arbitrary finite group, abelian or non-abelian.  Just like
GB codes can be seen as CSS codes constructed from a pair of index-two
quasicyclic codes, 2BGA codes are the smallest lifted-product (LP)
codes\cite{Panteleev-Kalachev-2020,Panteleev-Kalachev-2021}.

We give a formal expression based on idempotent matrices for the
dimension of general two-block codes.  The dimension is necessarily
even for such codes based on an abelian group
algebra\cite{Kalachev-Panteleev-2020} (which includes GB codes), as
well as for more general quantum two-block codes identified by certain
additional commutativity conditions.  We show that this constraint is
automatically satisfied for 2BGA codes based on a semi-simple group
algebra; the dimension of such codes is necessarily even.  This gives
a simple sufficient criterion for an essentially non-abelian 2BGA code
which cannot be reduced to such a code based on an abelian group.  We
also discuss the distance of 2BGA codes and, for a family of such
codes, give a lower bound in terms of distances of classical group
algebra codes.  In particular, this bound applies in the case where a
group algebra code reduces to an HP code.

The structure of the rest of the paper is as follows.  We introduce
necessary notations in Section \ref{sec:notations}.  Our main results
are given in Sec.~\ref{sec:2BGA}, followed by conclusions in
Sec.~\ref{sec:conclusions}.

\section{Preliminaries}
\label{sec:notations}

A classical code $\cal C$ linear over a finite field
$F\equiv \mathbb{F}_q$, where $q>1$ is a power of a prime $p$, the
characteristic of the field, with parameters $[n,k,d]_q$, is a $k$
dimensional vector space in $F^{n}$, the set of all length-$n$ strings
using elements of $F$ as characters.  Such a code can be specified in
terms of a generator matrix $G$ whose rows are vectors from ${\cal C}$
forming a complete basis, $\rank G=k$, or its parity-check matrix $H$
whose rows are orthogonal to any vector in $\cal C$, with $\rank H=n-k$,
\begin{equation}
  \label{eq:classical-code}
{\cal  C}={\cal C}_G\equiv{\cal C}_H^\perp,\quad GH^T=0.
\end{equation}
The codes ${\cal C}_G$ and ${\cal C}_H$ generated by rows of $G$ and
$H$, respectively, are called mutually dual.  The \emph{support} of a
vector $\bs x\equiv (x_1,x_2,\ldots,x_n)\in F^{n}$ is the set
of indices $i$ corresponding to non-zero components $x_i\neq0$, and
its Hamming weight is the size of the support.  The distance $d$ of a
linear code ${\cal C}$ is the smallest Hamming weight of a non-zero
vector in ${\cal C}$; by convention, $d=\infty$ for a trivial code
with $k=0$.

A very important class of codes are \emph{cyclic} linear
codes\cite{MS-book}, invariant under the group $C_n$ of cyclic
permutations.  A generalization to an arbitrary group are \emph{group
  codes}, or group algebra
codes\cite{Bazzi-Mitter-2006,Olteanu-vanGelder-2015,Milies-2019,%
  Borello-delaCruz-Willems-2022}.

Given a finite field $F$ and a finite group $G$, the group algebra
$F[G]$ is the linear space of all formal sums
\begin{equation}
  \label{eq:algebra-element}
  x\equiv \sum_{g\in G}x_g g,\quad x_g\in F,
\end{equation}
where group elements $g\in G$ serve as basis vectors,
equipped with the product naturally associated with the group
operation,
\begin{equation}
  \label{eq:FG-product}
  ab=\sum_{g\in G}\biggl(\sum_{h\in G} a_h b_{h^{-1}g}\biggr) g, \quad a,b\in F[G].
\end{equation}
Similar to cyclic codes, a left (right) group algebra code is
isomorphic to a left (right) \emph{ideal} $J$ in $F[G]$, defined as a
linear space such that for any $x\in J$ and any $r\in F[G]$, $rx\in J$
for the left ideal ($xr\in J$ for the right ideal).

The structure of ideals in $F[G]$ is particularly simple when
characteristic of the field and the group size are mutually prime,
$\gcd(p,|G|)=1$.  In this case, according to Maschke's theorem, the
group algebra is semisimple, and any ideal is a principal ideal
generated by an idempotent element, e.g., $J=e_J\cdot F[G]$ for a
right ideal $J=J_R$, with an idempotent $e_J^2=e_J\in J$ (see, e.g.,
Corollary 2.2.5 in Ref.~\onlinecite{Drozd-Kirichenko-book-1994}).

A quantum Calderbank-Shor-Steane (CSS)
code\cite{Calderbank-Shor-1996,Steane-1996} 
${\cal Q}\equiv\css(H_X,H_Z)$ can be defined as a direct sum of two
quotient spaces, $ {\cal Q}\cong {\cal Q}_X\oplus {\cal Q}_Z$,
\begin{equation}
  \label{eq:CSS}
 {\cal Q}_X={\cal C}_{H_Z}^\perp/{\cal C}_{H_X},\quad {\cal Q}_Z={\cal C}_{H_X}^\perp/{\cal C}_{H_Z}.
\end{equation}
For example, elements of ${\cal Q}_X$ are equivalence classes of
vectors in ${\cal C}_{H_Z}^\perp$, where two vectors are equivalent,
$x\simeq y$, if they differ by an element of ${\cal C}_{H_X}$,
$x-y\in {\cal C}_{H_X}$.  Such a pair of equivalent vectors are called
mutually \emph{degenerate}, while any vector in the equivalence class
of the zero vector is called \emph{trivial}.  The CSS \emph{generator
  matrices} $H_X$ and $H_Z$ have equal number of columns, $n$, and
orthogonal rows, $H_XH_Z^T=0$.  The parameters of the code
(\ref{eq:CSS}) are denoted $[[n,k,d]]_q$, where
\begin{equation}
  k=n-\rank H_X-\rank H_Z
  \label{eq:k-CSS}
\end{equation}
is the common dimension of the quotient spaces ${\cal Q}_X$ and ${\cal Q}_Z$, and
$d\equiv \min(d_X,d_Z)$ is the minimum weight of any non-trivial
vector in ${\cal Q}$, e.g.,
\begin{equation}
  \label{eq:d-CSS}
    d_Z\equiv d({\cal Q}_Z)=\min_{\bs u\in C_{H_X}^\perp\setminus C_{H_Z}}\wgt(\bs u).
\end{equation}
Physically, a quantum code $\css(H_X,H_Z)$ operates in a Hilbert space
${\cal H}_q^{\otimes n}$ associated with $n$ quantum-mechanical
systems of dimension $q$ each,
Galois-qudits\cite{eczoo-galois-qudits}, and a well defined
basis of $X$ and $Z$ operators acting in
${\cal H}_q^{\otimes
  n}$\cite{Ketkar-Klappenecker-Kumar-Sarvepalli-2006}.  Vectors of the
codes ${\cal C}_{H_X}$ and ${\cal C}_{H_Z}$ correspond to $X$- and
$Z$- operators in the stabilizer group whose generators must be
measured frequently during the operation of the code; generating
matrices $H_X$ and $H_Z$ with smaller row weights result in codes
which are easier to implement in practice.  Orthogonality condition
$H_XH_Z^T=0$ ensures that the stabilizer group is abelian.
Non-trivial vectors in ${\cal Q}_Z$ and ${\cal Q}_X$ correspond to $Z$
and $X$ logical operators, respectively.  Codes with larger distances
have logical operators which involve more qudits; such codes typically
give better protection against errors.

\section{Two-block codes}
\label{sec:2BGA}

In this work we discuss two-block CSS
codes with generator matrices
in the form\cite{Kovalev-Pryadko-Hyperbicycle-2013}
\begin{equation}
  \label{eq:css-blocks}
  H_X=(A,B),\quad H_Z^T={B\choose -A},
\end{equation}
where $A$ and $B$ are square commuting $\ell\times \ell$ matrices with
elements in $F$.  The commutativity guarantees the CSS
orthogonality condition, $H_XH_Z^T=0$.

{\bf Code dimension}: Given a square size-$\ell$ matrix $A$ with
elements in a finite field $F$, consider square idempotent matrices
$E_A$ and $F_A$ of the same size and rank such that
\begin{equation}
  \label{eq:idempotent-EA-FA}
  E_A^2=E_A,\quad F_A^2=F_A,\quad E_AA=AF_A=A.
\end{equation}
While these matrices are not unique, they can always be constructed
from the Smith normal form decomposition $A=U_AD_AV_A$, where $U_A$ and $V_A$
are square invertible matrices, and 
$D_A=\diag(1,\ldots,1,0,\ldots,0)$ has exactly $\rank A$ non-zero
elements along the diagonal.  Namely, we may choose
\begin{equation}
  E_A\equiv U_AD_AU_A^{-1},\quad  F_A\equiv V_A^{-1}D_AV_A.\label{eq:EA-FA-matrices}
\end{equation}

With idempotent matrices (\ref{eq:idempotent-EA-FA}), it is easy to
express the ranks of block matrices (\ref{eq:css-blocks}).  Indeed,
row and column transformations give (this is a simplified version of
more general expressions in Refs.~\onlinecite{Meyer-1970,Meyer-1973})
\begin{eqnarray}
  \nonumber 
  \rank H_X&=&\rank \left(
               \begin{array}[c]{cc}
                  A&E_AB\\ 0&(I-E_A)B
               \end{array} \right)\\   
           &=&\rank(A)+\rank (I-E_A)B, %%% \\
               \label{eq:rankHx}
\end{eqnarray}
and a similar result for the rank of the other matrix,
\begin{eqnarray}
%  \nonumber 
  \rank H_Z&=& \rank A+\rank B(I-F_A).
               \label{eq:rankHz} 
\end{eqnarray}
In general, $\rank H_Z\neq \rank H_X$.  However, the equality can be
achieved with some additional commutativity conditions. For example,
%%% if
%%% $E_A$ commutes with both $A$ and $B$, in which case we can take
%%% $F_A=E_A$, or
if both $E_A$ and $F_A$ commute with $B$, 
the second terms in the r.h.s.\ of Eqs.~(\ref{eq:rankHx}) and
(\ref{eq:rankHz}) are both equal $\rank B-\rank AB$.  This gives
\begin{statement} \label{th:rank-equal} Suppose that %%% {\em(i)}a
  %%% idempotent $E_A$ in Eq.~(\ref{eq:idempotent-EA-FA}) commutes with
  %%% both $A$ and $B$ in Eq.~(\ref{eq:css-blocks}), or {\em (ii)}
  idempotents $E_A$ and $F_A$ in Eq.~(\ref{eq:idempotent-EA-FA})
  commute with $B$ in Eq.~(\ref{eq:css-blocks}).  Then,
  \begin{equation}
  \label{eq:rank-equal}
  \rank H_X=\rank H_Z,\quad \text{and}\quad k=2(\ell-\rank H_X).
\end{equation}
\end{statement}
Evidently, Eq.~(\ref{eq:rank-equal}) also remains true after
interchanging the blocks, e.g., if  idempotents $E_B$ and $F_B$
commute with $A$.  

In particular, the conditions of Statement \ref{th:rank-equal} are
satisfied if $A$ has a square-free minimal polynomial.  Indeed, in
such a case $A$ can be diagonalized, $A=S^{-1}\Lambda S$, where square
matrix $S$ over $F$ is invertible, and the idempotents can be
constructed as $E_A=F_A=S^{-1} D S$, with $D$ a diagonal matrix with
elements equal to zero or one according to whether the corresponding
element of $\Lambda$ is zero or not.  It is easy to check that thus
constructed $E_A=F_A$ necessarily commute with $B$ if  $A$ does.

\textbf{Construction from classical group algebra codes}: To get a
pair of commuting matrices, we use an ansatz introduced by Panteleev
and Kalachev\cite{Panteleev-Kalachev-2020,Panteleev-Kalachev-2021}.
Namely, given an element $x\in F[G]$ of the group algebra with the
group size $\ell\equiv |G|$, the
$\ell\times\ell$ square matrices $\LL(x)$ and $\RR(x)$, respectively,
are defined by the left and right action on group elements,
\begin{equation}
  \label{eq:L-R-action}
  [\LL(x)]_{\alpha,\beta}\equiv \sum_{g\in G}x_g\delta_{\alpha,g\beta},\quad
  [\RR(x)]_{\alpha,\beta}\equiv \sum_{g\in G}x_g\delta_{\alpha,\beta g},
\end{equation}
where group elements $\alpha,\beta\in G$ are used to index rows and
columns, cf.~Eq.~(\ref{eq:algebra-element}), and
$\delta_{\alpha,\beta}=1$ if $\alpha=\beta$ and $0$ otherwise is the
Kronecker delta.  Row and column weights of $\LL(x)$ and $\RR(x)$ are
equal to $\wgt(x)$, the Hamming weight of the vector in $F^\ell$ with
components $x_\alpha$, $\alpha\in G$, which makes it easy to construct
sparse matrices.  Furthermore, for any $a,b\in F[G]$,
$\LL(a) \LL(b)=\LL(ab)$, $\RR(a)\RR(b)=\RR(ba)$, while a left and a
right matrices always commute,
\begin{equation}
\LL(a)\RR(b)=\RR(b)\LL(a).\label{eq:L-R-commute}
\end{equation}
With a group algebra element entirely supported on a subgroup,
$x_g\neq 0$ only if $g\in K<G$, one can also form smaller matrices,
e.g., $[\LL_K(x)]_{\alpha,\beta}$ of size $|K|\times |K|$, with
indices restricted to the same subgroup, $\alpha,\beta\in K$.  If we
introduce the {\em support group\/}\cite{Connell-1963}
\begin{equation}
  \label{eq:subgroup-Ga}
  G_x\equiv\left\langle\left\{ g\in G: x_g\neq0\right\}\right\rangle
\end{equation}
generated by elements of $G$ in the support of $x$, it is evident that
matrices $\LL(x)$ and $\RR(x)$ are block-diagonal (up to a
permutation), with square blocks of equal size $|G_x|$, corresponding
to, respectively, right and left cosets of the support group $G_x$
in $G$.  

With these definitions, the \emph{two-block group algebra} (2BGA)
codes, the CSS codes (\ref{eq:css-blocks}) with $A\equiv \LL(a)$
and $B\equiv \RR(b)$ given by Eq.~(\ref{eq:L-R-action}), are the
smallest lifted-product codes\cite{Panteleev-Kalachev-2021}
$\lp[a,b]$, where group algebra elements $a,b$ are treated as
$1\times 1$ matrices over $F[G]$.  Previously considered special cases
are GB codes\cite{Kovalev-Pryadko-Hyperbicycle-2013,%
  Panteleev-Kalachev-2019,Wang-Pryadko-2022}, with $G$ a cyclic group,
and \emph{abelian} 2BGA codes\cite{Kalachev-Panteleev-2020}, with $G$
an abelian group.

The structure of matrices $A$ and $B$ is such that the row labeled by
a group element $x\in G$ is associated, respectively, with the block
supported in the right coset $G_ax$ and that in the left coset $xG_b$.
When the product of the two support groups (the double coset
associated with the group identity element $1\in G$) does not contain
all group elements, $G_aG_b\subsetneq G$, the code $\lp[a,b]$ is
decomposed into smaller mutually disconnected subcodes associated with
different double cosets in $G_a\backslash G /G_b$.  The individual
\emph{double-coset} subcodes are not necessarily equivalent to each
other; it is well known that even the sizes of double cosets may
differ.

The case of GB codes\cite{Kovalev-Pryadko-Hyperbicycle-2013,%
  Panteleev-Kalachev-2019,Wang-Pryadko-2022} is recovered when $G$ is
a cyclic group,
$$
C_\ell\equiv \langle x\rangle\equiv \{1,x,x^2,\ldots, x^{\ell-1}\},
\quad x^\ell=1.
$$
There is an obvious one-to-one map between the group algebra
$F[C_\ell]$ and the ring of modular polynomials $F[x]/(x^\ell-1)$.
Then, a 2BGA code $\lp[a,b]$ is also a generalized-bicycle code
$\gb[a(x),b(x)]$ specified by a pair of polynomials $a(x)$,
$b(x)\in F[x]/(x^\ell-1)$, and the square blocks in
Eq.~(\ref{eq:css-blocks}) are just the circulant matrices $A=a(P)$ and
$B=b(P)$, where
\begin{equation} \label{eq:permutation-matrix} P =
  \begin{pmatrix}%eq2
    0&\ldots&0 &1\\
    1&&&0\\[-0.5em]
    &\ddots &&\vdots\\
    && 1&0
  \end{pmatrix}
\end{equation}
is an $\ell\times\ell$ cyclic permutation matrix.  A simple expression
for the dimension of a code $\gb[a,b]$ was given in
Ref.~\onlinecite{Panteleev-Kalachev-2019}.  In this case
$\rank H_X=\rank H_Z=\ell-\deg h(x)$, and
\begin{equation}%eq10
  \label{eq:GB-code-size}
  k=2\deg h(x),\quad   h(x)\equiv \gcd\left(a(x),b(x),x^\ell-1\right). 
\end{equation}
Evidently, Eq.~(\ref{eq:rank-equal}) is satisfied, as it also does
when the group $G$ is abelian\cite{Kalachev-Panteleev-2020}, or when
one of the subgroups, $G_a$ or $G_b$ [see Eq.~(\ref{eq:subgroup-Ga})],
is cyclic.  In the latter case the GB code is equivalent to a
quasi-cyclic LP code\cite{Panteleev-Kalachev-2020}.

More generally, consider \emph{semi-abelian} 2BGA codes satisfying the
conditions of Statement \ref{th:rank-equal}.  Namely, take a code
$\lp[a,b]$ where, e.g., $a\in F[G]$ is such that the corresponding right
$a\cdot F[G]$ and left $F[G]\cdot a$ ideals are generated by
idempotents $e_a$ and $f_a$, $e_aa=af_a=a$, and choose $E_A=\LL(e_a)$
and $F_A=\LL(f_a)$ to guarantee their commutativity with $B=\RR(b)$.
In particular, a semi-abelian 2BGA code is always obtained if the
group algebra $F[G]$ is semisimple.  Alternatively, we can select $a$
so that the corresponding subgroup $G_a$ in Eq.~(\ref{eq:subgroup-Ga})
has the order mutually prime with the field characteristic $p$,
$\gcd(p,|G_a|)=1$, so that only the subalgebra $F[G_a]$ be
semi-simple.  Then, the idempotents $e_a\in F[G_a]$ and
$f_a\in F[G_a]$ also generate the right and left ideals of $a$ in
$F[G]$, and, again, we can choose $E_A=\LL(e_a)$ and $F_A=\LL(f_a)$,
so that the conditions of Statement \ref{th:rank-equal} be satisfied.

To summarize, any abelian 2BGA code (including any GB code) or any
semi-abelian 2BGA code, e.g., based on a semisimple group algebra, has
an even dimension, see Eq.~(\ref{eq:rank-equal}).  Thus, any 2BGA code
with an odd dimension $k$ is \emph{essentially non-abelian}, i.e., it
is not permutation-equivalent to an abelian or a semi-abelian 2BGA
code.

\begin{example}
  Consider the alternating group $A_4$, also known as the rotation
  group of a regular tetrahedron,
  $$T=\langle x,y|x^3=(yx)^3=y^2=1\rangle,\quad |T|=12,$$ and the
  binary algebra $\mathbb{F}_2[T]$.  Select $a=1+x+y+x^{-1}yx$
  and $b=1+x+y+yx$ to get an \emph{essentially non-abelian} 2BGA code
  $\lp[a,b]$ with parameters $[[24,5,3]]_2$.
\end{example}

{\bf Distances of GB codes}: Several existence bounds for unrestricted GB
codes (without the limit on row weight) are given in
Ref.~\onlinecite{Wang-Pryadko-2022}.  In particular, with
$g(x)\equiv (x^\ell-1)/h(x)$ irreducible, cf.\
Eq.~(\ref{eq:GB-code-size}), a counting argument in the style of
Gilbert-Varshamov bound proves the existence of GB codes with $k=2$
and linear distance scaling (Example 8 in
Ref.~\onlinecite{Wang-Pryadko-2022}), and rate-$1/4$ GB codes with
$d\ge \sqrt\ell$ related to quadratic-residue cyclic codes (Example 9
in Ref.~\onlinecite{Wang-Pryadko-2022}).  This should be contrasted
with, e.g., HP codes whose distances satisfy the upper bound
$d<\sqrt n$.

In practice, we are more interested in quantum LDPC codes, with weight
of stabilizer generators not exceeding some fixed $w$.  Unfortunately,
the regular structure of GB codes is a disadvantage in this case, as
any such code is equivalent to a code local on a hypercubic lattice
$\mathbb{Z}^D$, with $D\le w-1$, or $D\le w-2$ if $\ell$ is prime
(Statement 13 from Ref.~\onlinecite{Wang-Pryadko-2022}).  With general
results from
Refs.~\onlinecite{Bravyi-Terhal-2009,Bravyi-Poulin-Terhal-2010}, this
gives upper bounds 
\begin{equation}
  \label{eq:finite-D-bounds}
  d\le \mathcal{O}(n^{1-1/D})\text{ and } kd^{2/(D-1)}\le \mathcal{O}(n). 
\end{equation}
Numerically, for a family of GB codes with $k=2$, the distance scaling
is consistent with $d=A(w)n^{1/2}+B(w)$, with $A(w)$ an increasing
function of $w$, although $d =\mathcal{O}(n^\alpha)$ with some
$\alpha= 1/2+\epsilon$ with a small $\epsilon>0$ cannot be
excluded\cite{Wang-Pryadko-2022}.

{\bf Lower distance bounds for 2BGA codes}: Best known are the usual
CSS bounds,
\begin{equation}
  \label{eq:lower-d-CSS-bound}
  d_Z\ge d(C_{H_X}^\perp), \quad d_X\ge d(C_{H_Z}^\perp).
\end{equation}
However, since the rows of $H_X$ and $H_Z$ are mutually orthogonal, we
have, e.g., $d(C_{H_X}^\perp)\le w_Z$, the minimum row weight of the
matrix $H_Z$.  Since our main interest is in highly-degenerate quantum
LDPC codes with bounded stabilizer weights and diverging distances,
the CSS bounds (\ref{eq:lower-d-CSS-bound}) are not very useful.

Consider the special case of a 2BGA code $\lp[a,b]$, with
$a,b\in F[G]$ such that the intersection subgroup
$N\equiv G_a\cap G_b$ is central in $G$.  In such a case, if we choose
two transversal sets of coset representatives, ${\cal A}$ from $G_a/N$
and ${\cal B}$ from $G_b/N$, any element of a double coset
$G_a\backslash x/G_b$ can be written as $\alpha x \beta\,\gamma$, with
$\alpha\in {\cal A}$, $\beta\in {\cal B}$, and $\gamma\in N$.  This
gives matrices $A$ and $B$ with individual square blocks of size
$c\equiv |N|$ given by, respectively, $\LL_N(a_{\alpha,\alpha'})$ and
$\RR_N(b_{\beta,\beta'})$, with matrix elements
$a_{\alpha,\alpha'},b_{\beta,\beta'}\in N$ defined by the action of
the two group algebra elements on the corresponding cosets, and
indices $\alpha,\alpha'\in {\cal A}$ and $\beta,\beta'\in {\cal B}$.
Explicitly, e.g., given the expansion (\ref{eq:algebra-element}) of
$a\in F[G]$,
$a_{\alpha',\alpha}=\sum_{\gamma\in N} a_{\alpha'\alpha^{-1}\gamma}
\gamma $.  This gives exactly the structure of a square-matrix
\emph{quasi-abelian} LP code\cite{Panteleev-Kalachev-2020} over the
group algebra $F[N]$, and also the following lower bound:

\begin{statement}[Version of Theorem 5 from
  \label{th:lower-d-central-intersection}
  Ref.~\onlinecite{Kovalev-Pryadko-Hyperbicycle-2013}] Given any two
  group algebra elements $a,b\in F[G]$ such that the intersection
  subgroup $N\equiv G_a\cap G_b$ of size $c\equiv |N|$ is central in
  $G$, consider classical codes with parity check matrices $A=\LL(a)$
  and $B=\RR(b)$.  Let
  $d_0=\min\left\{d(C_A^\perp),d(C_B^\perp)\right\}$ be the minimum of
  their distances.  Then, the distance $d_Z$ of the 2BGA code
  $\lp[a,b]$ satisfies the inequality $d_Z\ge \lceil d_0/c\rceil$.
\end{statement}

In fact, this lower bound becomes exact when the intersection subgroup
is trivial, $N=\{1\}$.  In this case each double-coset subcode of the
2BGA code $\lp[a,b]$ is equivalent to a hypergraph-product code
constructed from classical codes with parity-check matrices
$\LL_{G_a}(a)$ and $\RR_{G_b}(b)$ over the corresponding subgroups,
the individual blocks of $\LL(a)$ and $\RR(b)$.

It is known\cite{Borello-delaCruz-Willems-2022} that group algebra
codes include good codes with finite rates and finite relative
distances.  This guarantees the existence of finite-rate 2BGA codes
with distance scaling as a square root of block length.
Unfortunately, we do not have a matching upper bound for finite-rate
2BGA codes.

\section{Conclusions}
\label{sec:conclusions}

In conclusion, we considered a family of quantum two-block codes, an
ansatz particularly suitable for constructing short and
intermediate-length quantum LDPC codes.  This family includes
previously studied GB codes and their generalization, 2BGA codes,
which may be based on an abelian or a non-abelian group.  Compared to
``single-block'' quantum cyclic
codes\cite{Calderbank-1997,Thangaraj-McLaughlin-2001,%
  Aly-Klappenecker-Sarvepalli-2007} and a related construction based
on a general finite group\cite{Naghipour-Jafarizadeh-Shahmorad-2015},
the 2BGA codes have much more freedom: here the CSS orthogonality
constraint is naturally satisfied for any pair of group algebra
elements, and it is much easier to construct highly-degenerate quantum
LDPC codes.

We constructed a general expression relating the dimension of a
two-block code to those of single-block codes and, in the case of 2BGA
code $\lp[a,b]$, identified the cases of abelian, semi-abelian, and
non-abelian 2BGA codes, depending on the group $G$, the chosen group
algebra elements $a,b\in F[G]$, and the associated support groups
$G_a$ and $G_b$.  We also constructed a lower distance bound
applicable when the subgroup $N\equiv G_a\cap G_b$ is central in $G$.
The bound becomes exact when $N=\{1\}$, a trivial subgroup, in which
case the 2BGA code is equivalent to an HP code constructed from a pair
of group algebra codes.

Research in progress\cite{Lin-Pryadko-2023} includes enumeration of
2BGA codes with row weights $w\le 8$ for all inequivalent small groups
of size $\ell \le 50$.  Of particular interest are 2BGA codes with
larger $k$ which have many redundant minimum-weight stabilizer
generators and are expected to perform well in a fault-tolerant
setting as data-syndrome
codes\cite{Fujiwara-2014,Ashikhmin-Lai-Brun-2014,Ashikhmin-Lai-Brun-2016,%
  Zeng-Ashikhmin-Woolls-Pryadko-2019}.  This could enable single-shot
fault-tolerant quantum error
correction\cite{Bombin-2015,Campbell-2018}.  \medskip

\section*{Acknowledgments}
% \begin{acknowledgments}
We are grateful to Pavel Panteleev for helpful comments on an
early version of the manuscript.  This work was supported in part by the
APS M. Hildred Blewett Fellowship (HKL) and the NSF Division of
Physics via the grant 2112848 (LPP).
%\end{acknowledgments}

\bibliographystyle{IEEEtran}
\bibliography{IEEEabrv,lpp,qc_all,more_qc,ldpc,linalg,teach}

\end{document}